\documentclass[runningheads]{llncs}
\usepackage[T1]{fontenc}
\usepackage{graphicx}
\usepackage{booktabs}
\usepackage{tabularx}
\usepackage{ragged2e}
\usepackage{multirow}
\newcolumntype{Y}{>{\RaggedRight\arraybackslash}X}
\usepackage{pdflscape}

\begin{document}
\title{A Systematic AI Adoption Framework for \\ Higher Education: From Student GenAI Usage to Institutional Integration}
\titlerunning{A Systematic AI Adoption Framework for Higher Education}
%
\author{Michael Neumann\inst{1}\orcidID{0000-0002-4220-9641} \and
Lasse Bischof\inst{1}\orcidID{0009-0002-6622-0770}\and
Maria Rauschenberger\inst{2}\orcidID{0000-0001-5722-576X}\and
Eva-Maria Schön\inst{2}\orcidID{0000-0002-0410-9308}}
\authorrunning{M. Neumann et al.}
%
\institute{University of Applied Sciences and Arts Hannover,\\ Hannover, Germany 
\email{michael.neumann@hs-hannover.de}\\
\and
University of Applied Sciences Emden/Leer,\\  Emden/Leer, Germany
}
\maketitle              
\setcounter{footnote}{0}
\begin{abstract}
\textit{Context:} The rapid development of generative artificial intelligence technologies is transforming learning, assessment, and academic production in higher education. Despite increasing student adoption, many institutions lack operational mechanisms to systematically align regulations and curricula with evolving generative artificial intelligence practices, creating regulatory ambiguity and academic integrity risks.
\textit{Objective:} This study investigates how students utilize generative artificial intelligence tools in computer science-oriented disciplines and develops a structured, lightweight framework supporting institutional adaptation to pervasive GenAI usage.
\textit{Method:} We conducted a case study at the University of Applied Sciences and Arts Hannover (Germany), combining document analysis with an online survey (N = 151) targeting Business Information Systems and E-Government students. Quantitative responses were analyzed statistically, while open-ended responses underwent thematic synthesis.
\textit{Results:} Generative artificial intelligence adoption was widespread (85.43\%), with ChatGPT as the dominant tool. Students primarily used generative artificial intelligence for research assistance, programming support, and text processing. However, substantial policy uncertainty was observed: many students were unaware of or unsure about institutional generative artificial intelligence regulations. Document analysis revealed regulatory gaps, ambiguous terminology, and inconsistencies between formal rules and teaching practices.
\textit{Contribution:} To address these shortcomings, we propose the \textit{AI Adoption Framework for Higher Education}, an iterative and operational model integrating document analysis, empirical observation, synthesis of findings, and targeted updates of regulations and curricula. The framework addresses governance, assessment validity, and academic integrity under generative artificial intelligence conditions and provides practical guidance for institutional adaptation.


\keywords{Generative AI \and AI Adoption Framework \and Higher Education \and Academic integrity}
\end{abstract}
%

\section{Introduction}
\label{sec1:Intro}
The increasing availability and advancement of generative artificial intelligence (GenAI) have led to significant transformations across various domains, including higher education \cite{neumann_we_2023}\footnote{This paper is an extended version of our paper entitled: "``We Need To Analyze Students GenAI Use'':
Towards an AI Adoption Framework for Higher Education"~\cite{Bischof.2025}. This paper provides a detailed description of the AI Adoption Framework in Section~\ref{Sec5:Framework} and a discussion of the results providing practical implications~\ref{sec6:CallForAcademicIntegrity}.}. Among the different branches of AI, GenAI tools have gained interest due to their ability to generate human-like text, images, code, and other forms of content \cite{Freise.2025,Jimenez.2024,Ruedian.2025,Shailendra.2024}. 
These tools, such as ChatGPT by OpenAI, have introduced new opportunities and challenges in higher education and are available by the general public \cite{Grashoff.2024,Rasheed.2024,Sami.2025,Zhang.2024}, particularly in disciplines that involve substantial amounts of written assignments, research, and problem-solving \cite{Speth.2024}. 
In the context of higher education, students from diverse disciplines are increasingly integrating GenAI tools into their workflows \cite{von_garrel_kunstliche_2023}. 
The rapid adoption of these tools has sparked discussions regarding their potential benefits and risks in academia. 
While some educators perceive them as valuable aid in improving learning efficiency and fostering creativity, others raise concerns about ethical implications, the authenticity of academic work, and potential misuse of plagiarism or academic dishonesty~\cite{Chen.2020}.

Nowadays, higher education is challenged by several disruptive events of the last decade \cite{Schoen.2023}. In addition to the challenges due to the GenAI era outlined above, the Covid-19 pandemic is a good example. The widespread shift to online and/or distance learning during the Covid-19 pandemic has led to a major digital transformation, as universities around the world have adopted various tools for remote working and teaching, used virtual tutoring systems, or moved to cloud repositories that provide lecture materials over distance \cite{Matthies.2022,Neumann.2022}. 

Despite the pandemic, we see today new and upcoming challenges at an accelerating pace as we move towards an AI-driven era \cite{Foerster.2024,Schon.2023b}. Students with a stronger computer science focus like information science or e-government are leveraging these technologies for tasks such as requirements engineering \cite{Brockenbrough.2024}, code generation \cite{Maher.2023,Savelka.2023,Speth.2023}, report writing \cite{Datta.2024}, and summarizing complex concepts \cite{Speth.2024}. However, despite the growing use of GenAI tools, the extent to which students rely on these tools, as well as the nature of their specific applications, remains under-explored \cite{Zastudil.2023}. Additionally, institutional policies on AI usage in academia are still evolving, leading to uncertainties among students regarding acceptable AI-assisted practices in their studies \cite{Nithithanatchinnapat.2024}.\looseness=-1 

Given the profound potential of GenAI in higher education, it is crucial to examine how students utilize these tools \cite{neumann_we_2023}, identify the perceived benefits and challenges, and understand their implications for academic integrity as well as learning outcomes \cite{Chan.2023}.

In this paper, we aim to bridge the existing research gap by examining the adoption and impact of GenAI tools among students. Furthermore, we see a lack in developing the curricula of the study programs for the needed skills in the future. Thus, the above motivates the following research questions:

\begin{itemize}
    \item \textbf{RQ 1:} How do students utilize GenAI tools in their studies?\\ 
    With RQ 1, we gain an in-depth understanding of how students use GenAI tools in their studies.
    We focus strongly on the computer science-related fields of business information systems and E-Government in one German instituion to minimize biases of different academic fields and their particularities (\textit{e.g.,} skill-sets). 
    \item \textbf{RQ 2:} How can institutions systematically adapt their regulations and curricula with regard to AI through a structured framework?\\
    Using insights from RQ 1, we have created a framework to systematize our findings and provide universities with a tool to refine their curricula and adapt degree program regulations for AI usage.
\end{itemize}

Based on the findings of our case study, we present the main contribution of this paper: A framework to adopt GenAI into their study program curricula and regulations. To be precise in wording, we understand a \textit{framework} in this paper as a systematic described theoretical approach, which is in line with existing literature~\cite{Lederman.2015}.

The remainder of this paper is structured as follows: Section~\ref{Sec2:RelWork} reviews related work on GenAI in education, focusing on existing studies that have explored its adoption, benefits, and challenges. Section~\ref{sec3:ResearchDesign} outlines the research design, detailing the methodology, data collection process, and analytical approach. Section~\ref{sec4:Results} presents the key findings derived from the survey, followed by Section~\ref{Sec5:Framework}, which discusses the implications of the results for students, educators, and institutional policies. In Section~\ref{sec6:CallForAcademicIntegrity} we discuss the implications of our findings and present a call for academic integrity. Section~\ref{sec6:Limitation} addresses the study’s limitations, and Section~\ref{sec7:Conclusion} concludes the paper with a summary of findings and directions for future research.

\section{Related Work}
\label{Sec2:RelWork}
We searched for primary studies dealing with GenAI adoption in the Higher Educational Context.
First, we focused on existing frameworks for GenAI adoption. The \textit{4E framework}~\cite{Shailendra.2024} (Embrace, Enable, Experiment, Exploit) intends to integrate GenAI into higher education systematically. The iterative designed framework addresses curriculum design, roles of stakeholders, training requirements, ethical considerations, and evaluation mechanisms. Challenges such as academic integrity, privacy, and assessment fairness are also recognized. However, the framework has its limitations, especially in terms of detailed methodological guidance and empirical validation. Furthermore, there is a lack of explicit strategies for addressing potential technological biases or inaccuracies (\textit{e.g.,} data hallucination). 

Additionally, Su and Wang~\cite{Su.2023} introduced the \textit{IDEE framework} that consists of four steps. They include, \textit{e.g.,} outcomes identification, as well as an application example from practice. Though the framework is introduced on a level of detail, the practical application lacks on specific measures and actions.\looseness=-1 

Southworth et al. \cite{Southworth.2023} present the \textit{AI iteracy model} aiming to provide an opportunity for a systematic curriculum development of AI courses for
undergraduate study programs. The framework consists of five steps and is designed as an iterative approach. Even if this frameworks has its strengths in curriculum development, several other important facets such as the focus on regulations are lacking. Recent research further highlights that regulatory and policy-related aspects constitute a critical yet frequently underdeveloped dimension of GenAI integration. Rizki and Daoud~\cite{Rizki.2025} analysed institutional GenAI policies across eight New Zealand universities using Policy Discourse Analysis. Their findings indicate that while most institutions adopt an open but cautious stance towards GenAI, policy completeness and clarity vary considerably. The study emphasizes that fragmented or ambiguous regulations generate uncertainty among students and lecturers, potentially hindering consistent adoption practices. These results reinforce the argument that successful GenAI integration requires not only curricular or pedagogical frameworks but also coherent institutional governance structures.

To sump up, all of the identified models are multi-step frameworks, mostly applying an iterative approach aiming to provide a solution for the ongoing development of new or updated GenAI tools and technologies. However, all of the existing models are lacking on specific aspects such as artifacts and thus, do not provide an opportunity for us to apply them at our universities. Besides the strong focus on AI courses~\cite{Southworth.2023} the identified frameworks are mostly to detailed and heavyweight for an application on specific study programs. 
The main limitation of these frameworks is their failure to consider the specific circumstances at higher education institution. Specifically, they often overlook how students are currently using GenAI tools and the existing state of curricula and regulations. Consequently, we opted to develop our own data-driven framework.\looseness=-1

While existing adoption frameworks primarily emphasize curricular and structural considerations, empirical studies increasingly stress the role of human stakeholders. Mah and Groß~\cite{Mah.2024} investigated AI usage, self-efficacy, and professional development needs among faculty members in German higher education institutions. Their results reveal distinct usage profiles and indicate that lecturers’ AI self-efficacy significantly influences adoption behavior. Moreover, the study highlights that insufficient AI literacy and uncertainty regarding ethical and pedagogical implications remain central barriers. These findings suggest that GenAI integration cannot be conceptualized solely as a curricular or technological challenge but must incorporate the competencies, beliefs, and support needs of instructors.

In addition to the frameworks presented in the literature, we also found studies focusing on students' use of GenAI tools. Several empirical studies have examined how students engage with AI-driven tools, their frequency of use, and associated benefits and challenges. 

Beyond frequency-based analyses, recent qualitative research provides deeper insights into students’ actual interaction patterns with generative AI. Wang~\cite{Wang.2025} explored how students integrate ChatGPT into their writing processes using a phenomenological design. The study shows that learners employ GenAI across multiple stages, including brainstorming, structuring, revising, and linguistic refinement. Importantly, students reported tensions between leveraging AI support and maintaining their authentic voice, as well as concerns about potential learning loss. These findings suggest that GenAI usage is not merely a functional tool choice but reshapes cognitive and metacognitive learning processes, thereby reinforcing the need for empirically grounded integration strategies.

Research from von Garrel et al. \cite{von_garrel_kunstliche_2023} investigated the frequency and application of generative AI tools among university students across disciplines. The findings suggested that GenAI tools were predominantly used for text generation, summarization, and code development. The nationwide applied survey was conducted in Germany in 2023 and focused on student engagement with ChatGPT and similar GenAI tools. The study revealed that approximately 63.4\% of the surveyed students utilized generative AI tools for academic purposes, with significant variations depending on the field of study. Another study by Gottschling et al. \cite{gottschling_nutzung_2024} explored the implications of AI usage in student learning environments, highlighting the ethical concerns and academic integrity issues associated with AI-generated content. Beyond the German higher education context, recent large-scale empirical evidence further underlines the growing relevance of GenAI in universities. Aldossary et al. \cite{Aldossary.2024} examined students’ perceptions of generative AI tools across 15 Saudi universities (N = 1390). Their findings indicate a generally high level of awareness, acceptance, and perceived educational value of GenAI tools. Students particularly emphasized benefits such as improved understanding of complex concepts, enhanced feedback, and increased self-efficacy. At the same time, participants demonstrated awareness of potential challenges, including ethical concerns and misuse. These results reinforce the observation that students’ perceptions represent a critical factor for successful institutional adoption strategies and highlight the need for structured governance and usage policies.

Existing surveys aimed at thoroughly investigating and understanding students' application and use of GenAI tools are limited, particularly as they mostly include data from disciplines other than computer science. Therefore, we decided to design and conduct our own survey tailored specifically to this context. A detailed explanation of the applied research design is provided in the next section. Moreover, these studies have identified varying degrees of student awareness regarding institutional AI regulations. Some universities have established clear policies, whereas others are still in the process of defining guidelines. This discrepancy influences how students perceive and utilize AI tools in their coursework, underscoring the need for more structured institutional frameworks.

\section{Research Design}
\label{sec3:ResearchDesign}
Our research objectives focus on two main areas: First, the specific GenAI tools students use, their application areas, and how institutional guidelines influence their adoption. Second, we aim to develop a framework to facilitate adoption of AI within higher educational institutions.

\begin{figure*}[thb]
\centering
	\includegraphics[scale=0.34]{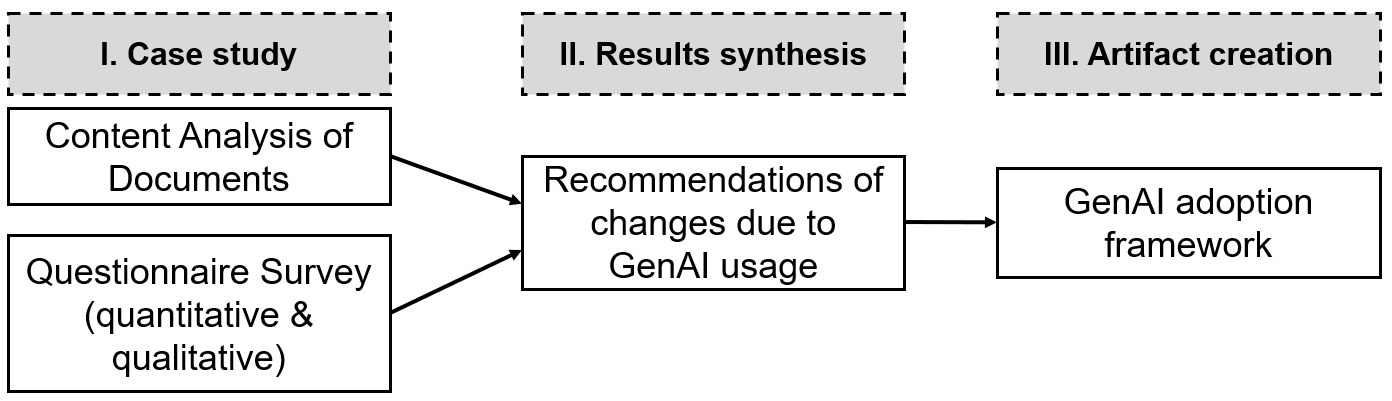}
	\caption{Research Design \cite{Bischof.2025}}
	\label{fig:researchdesign}
\end{figure*}

To address these research objectives, we designed a case study according to the guideline by Yin~\cite{Yin.2009} at the University of Applied Sciences and Arts Hannover focusing on the specific discipline of information systems. Figure~\ref{fig:researchdesign} depicts the research design including the applied research methods. 

\subsection{Case Context}
The University of Applied Sciences and Arts Hannover is located in Germany offering education in a wide span of disciplines such as design, engineering, economics, and computer science. In total, around 10,000 students are enrolled in the different programs in five faculties which are located over the city of Hannover. Our case study focus on the information systems discipline. The Department of Business Information Systems (Faculty IV) organizes and offers three different study programs for both a bachelor's and a master's degree. Here, we took the bachelor programs \textbf{B}usiness \textbf{I}nformation \textbf{S}ystems (BIS) and \textbf{E}-\textbf{Gov}ernment (EGOV) under study. In winter term 2024-2025 there were 465 of BIS students and 115 of EGOV students enrolled. A detailed overview per term is given in Table~\ref{tab:sample}.

The department is incorporating GenAI across different aspects of its study programs. Shortly after the launch of ChatGPT in November 2022, a working group in the department analyzed the opportunities and risks to adapt regulations and the impact on several exam types. Based on these analysis results, the university adopted specific GenAI related aspects to their regulations, especially for the bachelor thesis and  written exams. For instance, students have to respect guidelines for  theses and writing exams. These guidelines provide specific regulations related to the format (\textit{e.g.,} font, size, and line spacing), the citation style(s), and detailed information regarding the use of tools, including GenAI. By explicitly addressing GenAI, the guidelines aim to implicitly integrate this technology into the study programs. The last core adaption of the curriculum of both study programs dates back to 2018 and thus, the curricula do not define GenAI skills or further measures explicitly. However, in various courses in both study programs lecturers provide GenAI integration to the students. Examples are Requirements Engineering, Research Methods \& Scientific Writing, or Math.

\subsection{Data Collection \& Analysis}

\paragraph{Questionnaire Design:} We used an online survey as the primary data collection instrument. The choice of this method was driven by the need to obtain a broad and representative sample of students from the targeted disciplines at the case institution. The questionnaire was divided into four main parts. The questionnaire can be found at Zenodo~\cite{Bischof.2024}. The first part focused on demographic information, gathering details such as the study program. This information allowed for the categorization of responses and ensured that only relevant data from BIS and EGOV students were included in the analysis.

The second part explored students' attitudes toward GenAI tools. It assessed their familiarity with AI-based applications, their perceived usefulness, and their awareness of university policies concerning AI use. Responses in this part were recorded using a 5-point Likert scale, ranging from "\textit{strongly disagree}" to "\textit{strongly agree}", allowing for a nuanced understanding of student perspectives.

The third part investigated usage patterns and policy compliance, asking students whether they used GenAI tools and, if so, which ones. It also inquired about their primary use cases, such as assisting with research, programming, summarizing academic texts, or preparing for exams. Additionally, this part assessed whether students believed their AI usage aligned with university guidelines, which provided insight into potential gaps in institutional communication regarding AI regulations.

The final part covered perceived benefits and challenges associated with the use of GenAI in academic settings. While most of the questionnaire employed structured, multiple-choice questions, this part also included open-ended questions. This allowed students to elaborate on their experiences, including specific advantages and difficulties they encountered while integrating AI tools into their learning process.

\paragraph{Data Collection and Sampling Strategy:} Conducted between November 5, 2024, and December 31, 2024, the survey was hosted on LimeSurvey. Participation was entirely voluntary and anonymous to mitigate social desirability bias and ensure compliance with ethical standards related to student privacy.

To enhance the reliability of the data, the survey underwent a pilot phase involving 20 students from BIS and EGOV. Their feedback contributed to refining the questionnaire by improving clarity, ease of understanding, and the structure of the response format. Adjustments made after the pilot phase included rewording complex questions and modifying the layout for better readability.

The survey was distributed through multiple communication channels to maximize participation. Lecturers promoted the study in relevant courses, and QR codes linking to the survey were displayed in classrooms. Additionally, students received direct invitations via email and online university portals, ensuring that the target audience was adequately reached.

A non-probabilistic sampling approach was employed, meaning that students chose to participate voluntarily rather than being selected through a randomized process. While this method facilitated efficient data collection, it also introduced potential bias, as students with an interest in AI might have been more inclined to respond. 

\paragraph{Data Analysis:} The final data set ($N=151$) was reviewed and only includes fully completed submissions. 
The analysis was conducted using statistical software to ensure accurate processing. The dataset was cleaned and managed using Python’s Pandas library, while Matplotlib and Seaborn were used for visualizing trends. Further statistical computations, such as confidence intervals and significance testing, were performed using SciPy to verify the reliability of the findings.

The multiple-choice responses were analyzed quantitatively, with frequency distributions and comparative analyses conducted to identify patterns in GenAI tool adoption across different student groups. Open-ended responses were processed using a thematic analysis approach (according to \cite{Cruzes.2011}), in which recurring themes related to the benefits and challenges of AI adoption were categorized and interpreted.

\begin{table}[h]
    \centering
    \begin{tabular}{c|c|c|c}
        \toprule
        \textbf{Term} & \textbf{BIS} & \textbf{EGOV} & \textbf{Total Count} \\
        \midrule
        1 & 33 & 13 & 46 \\
        2 & 16 & 2 & 18 \\
        3 & 19 & 20 & 39 \\
        4 & 5 & 0 & 5 \\
        5 & 7 & 17 & 24 \\
        6 & 6 & 0 & 6 \\
        7 & 4 & 6 & 10 \\
        8 & 3 & 0 & 3 \\
        \midrule
        Overall Total & 93 & 58 & 151 \\
        \bottomrule
    \end{tabular}
    \caption{Number of students by term and study program~\cite{Bischof.2025}}
    \label{tab:sample}
\end{table}

\paragraph{Sample Description}
A total of 151 students participated in the survey, all of whom were enrolled at the case university. The sample consisted of 93 BIS (61.59\%) and 58 EGOV students (38.41\%). The sample distribution reflects the general student composition of these programs, although students in term 5 were underrepresented due to external internships during the survey period. Table~\ref{tab:sample} provides an overview of the students enrolled per term and study program.

In terms of academic level, most participants were in their first to fourth term, while students from the fifth and seventh terms were significantly less represented. This is attributed to their participation in off-campus practical training phases, which limited their engagement with university-related activities. However, keeping the total count of enrolled students in both programs in mind, we have a statistical significant sample for both programs (for BIS: 93 out of 465 or 20\%; for EGOV: 58 out of 115 or around 50\%).

\paragraph{Ethical Considerations:} To maintain ethical integrity, strict measures were implemented throughout the research process. Participation was fully voluntary, and students had the opportunity to withdraw at any stage without justifying. The survey was entirely anonymous, ensuring that no personal identifiers were collected. Additionally, formal approval from the university was obtained before data collection, ensuring that the study adhered to ethical research guidelines.

\paragraph{Synthesis and Framework Creation:} We analyzed questionnaire data alongside institutional documents to examine the use of GenAI tools, focusing particularly on students' use-cases and existing regulations governing their permitted usage.

Based on a thorough analysis of our datasets, we first synthesized our findings. We then discussed which recommendations for GenAI adoption could be identified. In a second step, considering the relationships among these recommendations, we developed a framework and elevated our approach for the selected case to a meta-level, aiming to provide a solution applicable to other academic institutions.

\section{Survey Results}
\label{sec4:Results}
Here, we answer our first research questions aiming to provide in-depth findings as a basis for the next section in which we will present our \textit{AI Adoption Framework for Higher Educational contexts}.

\subsection{GenAI Adoption by Students}
First, we present the answer of RQ 1: \textit{How do students utilize GenAI tools in their studies?}

One of the key findings of this study is the high adoption rate of GenAI tools among students. Of the 151 students surveyed, 129 students (85.43\%) reported that they actively use GenAI tools in their studies. Conversely, 19 students (12.58\%) indicated that they do not use these tools, while 3 students (1.99\%) abstained from answering. A breakdown of usage by study program reveals that 88.17\% of BIS students and 81.03\% of EGOV students use GenAI. This suggests that AI adoption is slightly higher among BIS students, possibly due to the stronger emphasis on computer science discipline and specific courses for programming and data-driven applications.

\paragraph{GenAI tools used by students:} Among the students who use GenAI, ChatGPT is by far the most frequently used tool, with 93.8\% of whom use it regularly. Other tools, such as Google Bard and DeepL Write, were mentioned but to a much lesser extent. The widespread preference for ChatGPT may be due to its intuitive user interface, availability, and broad range of functionalities. Additionally, students in computer science-related fields may favor ChatGPT over other AI tools due to its strong capabilities in code generation and debugging.

When asked about paid AI subscriptions, the vast majority of the AI-using respondents (85.27\%) reported that they rely exclusively on free versions of AI tools. Only 19 students (14.73\%) indicated that they use paid AI services, citing greater functionality and improved response accuracy as their primary reasons.

\paragraph{Use Cases of GenAI in academic work:} Students use GenAI tools for a variety of academic tasks, with the most common applications being:

\begin{itemize}
    \item  Research Assistance (40.15\%): AI is frequently used to summarize academic papers, generate explanations, and structure research topics.
    \item Programming Support (37.8\%): A large percentage of students, particularly from BIS, use GenAI for coding help, debugging, and generating sample code snippets.
    \item Text Summarization (34.65\%): Many students rely on GenAI tools to condense lecture notes, academic papers, and textbooks into more digestible formats.
\end{itemize}

Other reported applications include exam preparation, language translation, and writing assistance for essays and reports (also known as written exams). However, students also acknowledged that over-reliance on GenAI for academic tasks can limit critical thinking skills and independent problem-solving.

\paragraph{Perceived challenges of GenAI usage by students:} A major concern revealed in the survey is the lack of clarity surrounding institutional regulations on AI usage. 


The majority of students, regardless of most academic terms (with the exception of term 6), report that existing regulations are unknown to them. We initially expected that the number of students familiar with the regulations would steadily increase with each semester, while the number of those unaware of them would decrease accordingly. However, the data suggest that a significant degree of uncertainty regarding existing rules and regulations persists until the end of the study program. 
Consequently, we investigated the extent to which students believe their use of GenAI adheres to established guidelines and regulations:
\begin{itemize}
    \item 43.7\% of students were unsure whether their AI usage was compliant with academic policies.
    \item 37.2\% believed their use was aligned with the university’s rules.
    \item 19.1\% admitted they might be violating institutional policies.
\end{itemize}

This high degree of uncertainty highlights a gap in institutional communication, as many students reported a lack of clear AI usage policies from faculty members and university administration. Nevertheless, based on our document screening, we found that the internal university communication rules imply mentioning existing guidelines and regulations. Thus, we also see a lack of specific skills (or motivation) on the student side to verify the existing guidelines for GenAI usage rules. However, one may assume that some students may unknowingly violate academic integrity standards due to unclear or inconsistent enforcement. Furthermore, concerns about over-reliance on GenAI tools were also raised. A significant portion of students warned that relying too heavily on AI can impede their ability to develop a deeper understanding, diminish critical thinking skills, and promote passive learning habits. Furthermore, 12 students explicitly mentioned concerns related to data privacy, particularly regarding the storage and use of personal input data by AI companies. 


In conclusion of our findings on the use of GenAI by students, we compared our findings to the results from the existing literature. Here, we focused our comparison to a previous nationwide survey(\cite{von_garrel_kunstliche_2023}), which reported that only 63.4\% of students in Germany used GenAI. In contrast, our study found a significantly higher GenAI usage rate of 85.43\%. A Chi-square test confirmed that this difference is statistically significant ($\chi^2$ = 30.09, p < $4.1 \times 10^{-5}$). However, the survey (\cite{von_garrel_kunstliche_2023}) was conducted in 2023 and considered a wide spread of study programs, while ours focused strongly on information systems and related disciplines. Thus, the results discrepancy may be due to the increasing prevalence of GenAI tools over time, or the higher affinity of business informatics and e-government students for GenAI technologies compared to students from other disciplines. Nevertheless, especially universities providing study programs with a relation to computer science, information systems or similar should be aware that GenAI adoption by the mass of their students may be the reality; for our case university it is. Thus, universities should aim to integrate AI literacy into academic programs, ensure clear policy communication, and encourage responsible AI use. By doing so, they can prepare students for an increasingly AI-driven world while preserving the integrity of higher education. These observations indicate that institutional responses to GenAI often remain reactive and fragmented. While regulatory adaptations represent an important first step, they do not fully address the pedagogical and competency-related implications of pervasive GenAI usage. 

The necessity of a structured adoption framework is further supported by findings from adjacent educational research. Systematic analyses of AI ethics education demonstrate that higher education institutions frequently struggle to align pedagogical innovation with robust assessment and competency development mechanisms~\cite{Wiese.2025}. Given that GenAI technologies significantly amplify ethical and epistemic challenges, adoption strategies that focus solely on regulatory adjustments risk remaining incomplete. Instead, institutional adaptation requires coordinated consideration of governance, curricular design, and evaluation practices, which our framework explicitly addresses. 

In response to these challenges, we propose the AI Adoption Framework for Higher Education, which is presented in the following section.

\section{AI Adoption Framework for Higher Education Organization}
\label{Sec5:Framework}
The findings of this study underscore the transformative impact of GenAI on higher education. As GenAI tools become increasingly sophisticated, educational institutions must strike a balance between leveraging AI’s benefits and maintaining academic integrity. Our case study results suggest that while students readily adopt GenAI, institutional policies have not yet fully adapted to this technological shift. Moving forward, a proactive and informed approach is essential. Based on a thorough analysis of our findings, we answer our second research question: \textit{How can institutions systematically adapt their regulations and curricula with regard to GenAI through a structured framework?}

\subsection{Design Principles for the Framework}
\begin{figure*}[thb]
\centering
	\includegraphics[scale=0.32]{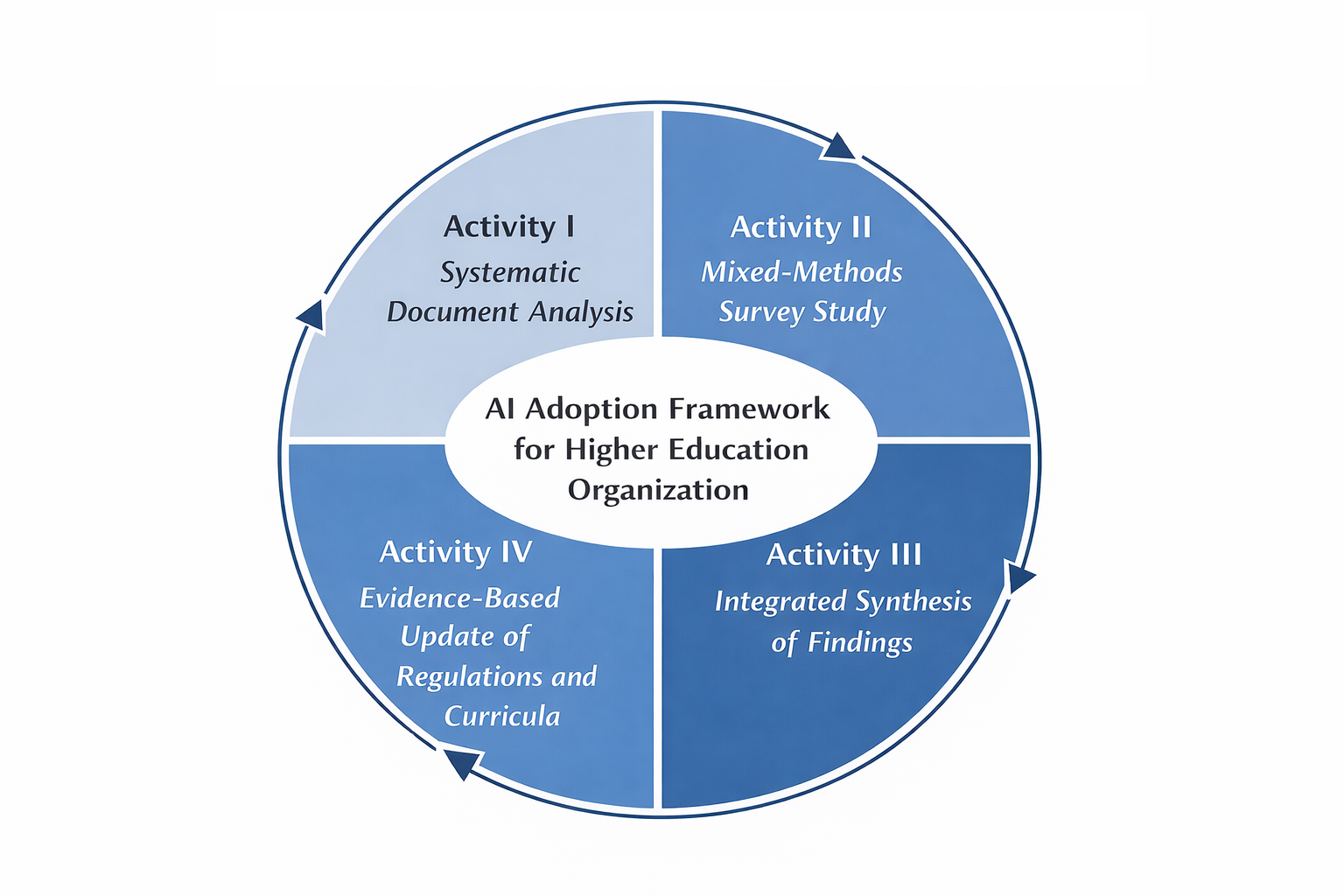}
	\caption{Visualization of the AI Adoption Framework}
	\label{fig:AdoptionFramework}
\end{figure*}

The development of our AI Adoption Framework was guided by several practical and methodological considerations derived from both our empirical findings and the institutional context of higher education. Rather than proposing a comprehensive or heavyweight governance model, we intentionally designed a framework that can be applied with limited resources and within existing academic structures in an iterative manner. Figure~\ref{fig:AdoptionFramework} depicts the framework.

Higher education institutions operate in highly regulated environments where curricular changes, examination regulations, and policy adaptations typically require coordination across multiple committees and organizational units. Consequently, any framework intended for institutional use should remain lightweight, transparent, and compatible with established decision-making processes.

Our survey findings indicate that student adoption of GenAI tools is already widespread, evolving rapidly, and in terms of the art of usage volatile. The use of todays tools may change to other tools consistently. This dynamic usage behavior requires an approach that does not assume stable technological conditions. Instead, the framework must accommodate continuous technological developments, including the emergence of new GenAI models, changing capabilities, and updated legal or ethical guidelines. Existing frameworks for AI or GenAI adoption often focus on curriculum design or teaching strategies in isolation (see Section~\ref{Sec2:RelWork}). However, our document analysis and empirical results highlight that regulations, guidelines, and assessment rules are equally critical factors influencing GenAI integration. Therefore, our framework explicitly connects regulatory adaptation with curricular considerations.

Finally, we emphasize an evidence-informed perspective. Institutional responses to GenAI should not be driven solely by assumptions, anecdotal experiences, or external pressures. Instead, they should be grounded in systematic observations of actual student practices and the current state of institutional documents. These considerations collectively shaped the structure and iterative logic of the proposed framework.

\subsection{Framework Steps}
To support practical application, the four activities of the AI Adoption Framework can be understood as complementary rather than strictly sequential steps or phases. Below, we explain the four activities on detail and provide specific examples for each activity.

\paragraph{Activity I - Document Analysis:} The document analysis activity establishes a structured understanding of the institutional context in which GenAI adoption occurs. Rather than performing an abstract policy review, this activity may be implemented as a systematic screening of regulatory and curricular artifacts that govern academic practices.

In practical settings, institutions typically begin by identifying relevant document types, including examination regulations, thesis guidelines, module/curricula descriptions, and academic policies. A structured analysis may then assess whether these documents explicitly address GenAI technologies, contain ambiguous terminology, or omit guidance entirely.

For example, a bachhelor/master thesis guideline may require students to produce “independent work” without clarifying whether AI-assisted content generation is permissible. In such cases, the document does not necessarily prohibit GenAI usage but introduces interpretative uncertainty. Similarly, examination regulations may prohibit “unauthorized aids” without specifying whether GenAI tools fall under this definition. Table~\ref{tab:activity1_documents} provides an overview of possible document types and analyzation tasks.

\begin{table}[t]
\raggedright
\caption{Practical examples of Document Analysis in Activity I}
\label{tab:activity1_documents}
\begin{tabularx}{\linewidth}{@{}p{3.4cm}p{3.3cm}Y@{}}
\toprule
\textbf{Document type} & \textbf{Typical examples} & \textbf{Details} \\
\midrule

Examination regulations &
Program examination regulations; faculty-level exam rules &
Whether GenAI qualifies as an aid; permitted vs. prohibited tool usage; definitions of misconduct; consistency with assessment formats. \\

Academic integrity / misconduct policy &
Code of conduct; plagiarism policy; integrity guidelines &
Definitions of independent work and originality; treatment of AI-generated content; disclosure obligations; enforcement logic. \\

Thesis guidelines &
Bachelor/Master thesis guide; writing guidelines &
Permissibility of GenAI tools; disclosure requirements; acceptable assistance boundaries; responsibility statements. \\

Course/module descriptions &
Learning outcomes; module handbook &
Presence of GenAI-related competencies; alignment between learning objectives and GenAI capabilities; assessment implications. \\

Assessment-specific guidance &
Written exam rules; take-home exam instructions &
Allowed support tools by assessment type; clarity of boundaries; documentation requirements. \\

Data protection / IT policies &
Privacy policy; IT usage policy; tool approval lists &
Approval status of GenAI tools; restrictions on sensitive data; rules for external AI systems. \\

Teaching guidelines / didactic recommendations &
Internal teaching handbook; AI usage guidelines &
Consistency with regulations; communication of GenAI policies; recommended instructional practices. \\

\bottomrule
\end{tabularx}
\end{table}

Document analysis may also reveal inconsistencies across artifacts. During our case study, informal teaching practices in certain courses encouraged GenAI usage for tasks such as code generation or text refinement, while formal regulations remained silent on AI-assisted workflows. Identifying these discrepancies constitutes a central outcome of this activity, as they indicate areas requiring clarification rather than immediate restriction. We prepared a checklist as a basis for identifying these discrepancies (see Table~\ref{tab:activity1_checklist} below). 

\begin{table}[t]
\raggedright
\caption{Specific analysis checklist for Activity I with example indicators.}
\label{tab:activity1_checklist}
\begin{tabularx}{\linewidth}{@{}p{3.3cm}Yp{3.1cm}@{}}
\toprule
\textbf{Checklist item} & \textbf{How to assess it (practical)} & \textbf{Example indicator} \\
\midrule

Explicit GenAI mention &
Screen documents for references to AI, GenAI, or tool usage &
No explicit AI reference in examination regulations. \\

Definition clarity &
Review how key terms (e.g., independent work, aids) are defined &
``Independent work'' remains unspecified. \\

Permitted vs. non-permitted uses &
Identify whether acceptable GenAI practices are distinguished &
Language correction allowed; content generation unclear. \\

Disclosure requirements &
Check for mandatory disclosure or attribution mechanisms &
No GenAI disclosure statement defined. \\

Assessment alignment &
Compare assessment formats with GenAI capabilities &
Take-home tasks easily supported by GenAI tools. \\

Consistency across documents &
Cross-check for contradictions or regulatory gaps &
Course-level policies conflict with exam rules. \\

Enforcement responsibility &
Examine whether responsibilities and procedures are defined &
No clear responsibility for AI-related violations. \\

Data protection constraints &
Review restrictions on sensitive data and external AI tools &
Lack of GenAI-specific data usage guidance. \\

Update mechanism &
Check for review cycles or explicit policy ownership &
Regulations predate widespread GenAI adoption. \\

\bottomrule
\end{tabularx}
\end{table}

\paragraph{Activity II - Quantitative and Qualitative Surveys:} While document analysis captures institutional assumptions, surveys provide insights into actual student practices. This activity acknowledges that GenAI adoption is largely driven by students’ everyday academic workflows and may therefore diverge from regulatory expectations.

To generate decision-relevant results, surveys may connect usage patterns with policy awareness and perceived compliance. For instance, students may be asked which GenAI tools they use, for which academic tasks, and whether they are aware of institutional regulations governing AI usage.

Our findings illustrate the necessity of this approach. Although a substantial majority of students reported active GenAI usage, many expressed uncertainty regarding whether their practices aligned with university policies. A typical response scenario involved students using GenAI tools for summarizing lecture materials or assisting with programming tasks while being unsure whether such usage required disclosure or was restricted. This uncertainty is particularly relevant as uploading copyrighted materials into AI systems without appropriate authorization may raise legal and ethical concerns, especially regarding intellectual property and usage rights.

Qualitative responses further contextualize these patterns. Students may describe situations in which instructors explicitly allowed GenAI usage for brainstorming or code debugging, contrasted with assessment contexts where rules remained unclear. Such observations highlight that regulatory uncertainty often emerges not from deliberate misconduct but from ambiguous or inconsistently communicated expectations.

\paragraph{Activity III - Synthesis of Findings:} The synthesis activity consolidates institutional and empirical observations into actionable interpretations. Rather than presenting findings in isolation, this activity connects document-level gaps with student-level experiences.

For example, if document analysis identifies missing references to GenAI tools while surveys indicate widespread adoption, synthesis may interpret this discrepancy as regulatory lag. Similarly, high levels of student uncertainty regarding compliance may suggest deficiencies in policy communication rather than systematic violations.

In practical terms, synthesis may result in structured recommendations. A combination of high GenAI usage and low policy awareness may lead to proposals for explicit disclosure mechanisms. Observed inconsistencies between module descriptions and student practices may motivate targeted curriculum updates.

Importantly, synthesis also reframes potential conflicts. Divergences between regulatory language and student interpretations should not automatically be treated as compliance failures. Instead, they may reflect the rapid emergence of technologies that were not considered during the original formulation of institutional policies.

\paragraph{Activity IV - Updating Regulations and Curricula:} The updating activity translates synthesized findings into institutional adaptations. Given the complexity of academic governance processes, incremental adjustments often represent the most feasible strategy.

For instance, thesis guidelines may be revised to explicitly acknowledge GenAI usage. A previously generic requirement for “independent work” may be reformulated to clarify that GenAI tools are permissible as supportive instruments provided that their usage is transparently disclosed. Such a modification reduces interpretative ambiguity without imposing blanket prohibitions. Furthermore, this transparency makes it easier for the examiner to assess the student's personal contribution.

Similarly, institutions may introduce lightweight disclosure statements. Students may be required to include a brief declaration specifying whether GenAI tools were used and for which purposes. This mechanism enhances transparency while preserving flexibility across disciplines and assessment formats.

Curricular adaptations may follow analogous principles. If surveys reveal extensive GenAI reliance combined with limited critical reflection skills, targeted learning objectives may be introduced. For example, modules may incorporate components addressing the verification and evaluation of AI-generated outputs rather than focusing solely on tool usage.

Assessment practices may also be reconsidered. In contexts where GenAI tools significantly alter task execution, complementary evaluation formats such as oral examinations or reflective reports may preserve academic integrity while acknowledging technological realities.

\subsection{Iterative Adaptation and Institutional Learning}
The iterative structure of the framework reflects the dynamic nature of GenAI technologies. Institutional policies and curricular structures developed under pre-GenAI assumptions may rapidly lose alignment with emerging student practices.

In practical application, early iterations may prioritize regulatory transparency and clarification. Subsequent cycles may address curriculum integration, assessment design, or communication strategies. For example, an institution may initially introduce disclosure rules, later evaluate student comprehension of these policies, and refine guidance based on observed challenges.

Iteration therefore functions as a mechanism for continuous institutional learning rather than repetitive policy revision. This perspective acknowledges that GenAI adoption is not a transient phenomenon but an evolving condition requiring sustained adaptation.

\section{Discussion and Call for Academic Integrity}
\label{sec6:CallForAcademicIntegrity}
\subsection{Practical Implications}
The results of our study and the AI Adoption Framework should be interpreted within the broader context of existing research on GenAI adoption and AI integration in higher education. Prior studies and conceptual frameworks have already emphasized that generative AI technologies introduce both opportunities and structural challenges for educational institutions. However, our findings highlight a dimension that remains comparatively underexplored: the interaction between institutional regulations, student practices, and academic integrity under conditions of widespread GenAI usage.

Several frameworks discussed in the related work provide valuable perspectives on AI adoption. For instance, the 4E framework (Embrace, Enable, Experiment, Exploit)~\cite{Shailendra.2024} proposes an iterative model for integrating GenAI into higher education, emphasizing curriculum design, stakeholder roles, and evaluation mechanisms. Similarly, the IDEE framework~\cite{Su.2023} outlines structured steps for educational AI integration, while AI literacy models focus on competency development and curriculum transformation. These models share a common assumption: institutions must proactively engage with GenAI rather than treating it as a peripheral phenomenon.

Our empirical findings reinforce this assumption but also extend it. While existing frameworks primarily address curriculum design, pedagogical innovation, or competency development, our case study reveals that regulatory ambiguity and policy awareness constitute equally critical factors. A substantial proportion of students reported uncertainty regarding compliance with institutional rules, despite the presence of formal regulations. This observation suggests that the effectiveness of governance structures cannot be evaluated solely by their existence but must consider their interpretability and alignment with actual practices.

Furthermore, our results complement earlier empirical studies on GenAI usage. For example, prior surveys conducted in Germany~\cite{von_garrel_kunstliche_2023} reported significant but lower adoption rates of GenAI tools. The markedly higher usage rates observed in our study may reflect both the rapid technological diffusion since earlier investigations and the disciplinary context of information systems. More importantly, however, the persistence of policy uncertainty across academic terms indicates that institutional adaptation challenges are not automatically resolved through increased student experience or exposure.

When contrasted with existing adoption frameworks, our AI Adoption Framework (see Section~\ref{Sec5:Framework}) emphasizes a distinct operational focus. Rather than centering exclusively on curriculum transformation or AI literacy, the framework explicitly integrates document analysis, empirical observation, and iterative regulatory updates. This orientation reflects the empirical reality that GenAI adoption is not merely a pedagogical issue but a governance and integrity challenge.

The practical implications emerging from this combined perspective are as follows:
First, institutions should acknowledge that GenAI usage is likely to persist regardless of restrictive or undefined policies. Earlier conceptual models already stress the need for iterative adaptation, yet our findings demonstrate that misalignment between regulations and student practices generates uncertainty rather than compliance. In the consequence, context-sensitive guidance is therefore not optional but necessary.

Second, assessment practices require systematic reconsideration. AI literacy frameworks and curriculum models emphasize competency development, but GenAI technologies directly affect the validity of traditional assessment formats. Without adaptive strategies, institutions risk evaluating students’ ability to effectively leverage AI systems rather than their conceptual understanding or methodological competence. However, mere operational or application-oriented proficiency in using GenAI tools does not adequately reflect deeper disciplinary knowledge, critical reasoning abilities, or the capacity to independently construct and evaluate solutions.

Third, regulatory communication must be treated as a core governance function. Prior literature (e.g., \cite{Chen.2020,gottschling_nutzung_2024} discusses ethical considerations and policy development, yet our data indicate that awareness and interpretation gaps remain widespread. Institutional integrity structures depend not only on rule formulation but also on their consistent articulation and shared understanding.

These observations converge on a broader concern that resonates with existing research: the preservation of academic integrity. Academic integrity frameworks historically rely on assumptions of authorship, originality, and demonstrable competence. GenAI technologies challenge these assumptions by enabling new forms of assistance that blur traditional distinctions between tool usage and intellectual contribution.


\subsection{Call for Academic Integrity}
The emergence of GenAI technologies requires higher education institutions and academic stakeholders to re-examine the operational foundations of academic integrity. Existing frameworks for AI adoption, curriculum design, and AI literacy underscore the need for proactive institutional engagement. Building on these perspectives and our empirical observations, we articulate the following propositions calling higher education institutions, researchers, and lecturers for preserve academic integrity:

\paragraph{Proposition 1 - Academic Integrity Requires Explicit Normative Boundaries for AI-assisted Work:} The increasing integration of GenAI tools into academic workflows necessitates explicit normative boundaries that define acceptable and non-acceptable forms of AI assistance. Traditional academic conventions implicitly assume a clear distinction between tool usage and intellectual contribution. GenAI systems blur this distinction by enabling content generation, transformation, and reasoning support at unprecedented scales.

Our findings indicate that students frequently operate under conditions of interpretative uncertainty, particularly when institutional regulations rely on generic terminology. In the absence of clearly articulated boundaries, individuals are required to infer permissible practices, leading to inconsistent interpretations across courses and assessment contexts.

Institutions should therefore establish explicit normative reference frameworks that specify the role of GenAI tools in learning, assessment, and academic production. Such frameworks must not remain abstract value statements but should operationalize categories of permissible assistance, disclosure obligations, and responsibility structures. Without this clarification, academic integrity risks becoming situationally negotiated rather than institutionally defined.

\paragraph{Proposition 2 - The Credibility of Academic Qualifications Depends on Human-Centered Evaluation:} Academic qualifications derive their epistemic and societal legitimacy from the assumption that demonstrated achievements reflect individual understanding, reasoning, and accountability. While GenAI systems may support learning processes, the delegation of evaluative judgment to automated mechanisms introduces fundamental risks to the interpretability and validity of academic assessment.

Unlike administrative or analytical tasks, academic evaluation involves contextual interpretation, normative reasoning, and disciplinary judgment that cannot be fully reduced to computational outputs. Over-reliance on automated evaluation mechanisms may therefore undermine the reliability and defensibility of assessment decisions.

Maintaining human-centered evaluation is not merely a pedagogical preference but a structural requirement for preserving the credibility of academic degrees. Where assessment processes lose their human interpretative core, the signaling function of academic qualifications for both within academia and in broader societal contexts may erode. In this context, however, it is important to be transparent about the use of AI tools and to document this use so that examiners can assess the candidate's individual contribution.

\paragraph{Proposition 3 - GenAI Amplifies the Need for Methodological Transparency and Verifiability:} The proliferation of GenAI tools intensifies longstanding challenges related to the verification, reproducibility, and interpretability of academic work. GenAI-generated outputs may appear plausible while lacking epistemic grounding, thereby increasing the difficulty of assessing reliability through conventional means.

Academic integrity cannot be sustained solely through restrictions on tool usage. Instead, it requires a stronger emphasis on methodological transparency, documentation practices, and verifiability of knowledge production processes. Open Science principles, including reproducibility, traceability, and explicit methodological articulation, provide structural mechanisms for mitigating risks associated with opaque or unverifiable AI-assisted contributions.

In GenAI-mediated environments, the absence of transparent methodological practices may lead to a gradual erosion of trust, as evaluators and peers face increasing difficulty in distinguishing between reasoned analysis and automated generation. Reinforcing methodological competence and transparency therefore represents a necessary adaptation to maintain epistemic robustness.

\paragraph{Proposition 4 - Institutional Integrity Requires Explicit, Enforceable, and Observable Rules:} Academic integrity cannot be sustained through implicit expectations or abstract value statements alone. Institutions must establish explicit, operationalizable, and enforceable rules governing GenAI usage across assessment, coursework, and academic writing contexts. Our findings indicate that regulatory ambiguity is a primary source of uncertainty, which in turn increases the risk of inconsistent practices and unintentional violations.

From an institutional perspective, regulations must move beyond generic formulations such as “independent work” or “unauthorized aids,” as these terms no longer provide sufficient guidance in GenAI-mediated environments. Instead, institutions should define concrete categories of permissible and non-permissible AI-assisted practices, disclosure requirements, and accountability mechanisms. Nevertheless, we have to consider that with customized regulations the complexity from students perspective arise. While generic regulations have the advantage that they build a common ground for students to provide information of what is allowed or what is forbidden, customized guidelines by lecturers may lead to the situation that the students have to adjust to several different regulations during their studies. This challenge can be solved in providing both a generic regulation ground and a synthesized customization potential, which can be ensured by an established understanding of the appropriate degree of customization among the lecturers, courses, and assessments. 

Furthermore, integrity structures must be observable in practice. Rules that are formally documented but not systematically communicated, monitored, or embedded in assessment workflows have limited effect. Sustainable academic integrity therefore depends not only on policy formulation but on institutional mechanisms ensuring that expectations are interpretable, consistently applied, and regularly revisited.

\paragraph{Proposition 5 - Academic Integrity Depends on Designing Assessments That Remain Meaningful Under GenAI Conditions:} The preservation of academic integrity is inseparable from assessment design. GenAI technologies fundamentally alter how academic tasks can be executed, thereby challenging traditional assumptions about what constitutes evidence of competence. Integrity cannot be defended solely through detection mechanisms or prohibitive policies; it must be embedded in assessment structures that remain meaningful despite pervasive AI assistance.

Institutions and educators should therefore critically examine whether existing assessment formats continue to measure the intended competencies. Tasks that can be largely externalized to GenAI systems without requiring demonstrable understanding may undermine the validity of evaluation processes. In such contexts, adaptive strategies including process-oriented evaluation, oral components, or reflective elements become necessary.

This proposition does not imply rejecting GenAI technologies. Rather, it emphasizes that academic integrity requires alignment between technological realities and evaluation logic. Degrees and certifications derive their societal value from the credibility of demonstrated competencies. Where assessment designs fail to reflect GenAI capabilities, this credibility may erode.

\section{Limitations}
\label{sec6:Limitation}
As detailed in Section~\ref{sec3:ResearchDesign}, we employed a systematic approach to prepare and conduct our case study. However, all studies have limitations. In this section, we address these limitations using the validity threats concept outlined by Runeson and Hoest~\cite{Runeson.2009}.

\paragraph{Construct Validity} The survey questions may have introduced interpretation bias, particularly regarding terms like "\textit{academic integrity}" and "\textit{rule compliance}", due to the absence of clear higher education guidelines on GenAI usage. Additionally, self-reported GenAI usage may be inaccurate due to social desirability bias, despite anonymity. The framing of questions, including specific GenAI tool examples, may have led to an anchoring effect, influencing respondents' selections rather than capturing their broader GenAI usage patterns.

\paragraph{Internal Validity} As this study relies on cross-sectional survey data, it identifies associations rather than causal effects. A key confounding variable is students' prior technical expertise, as those with programming or data science backgrounds may be more inclined to use GenAI tools, potentially skewing results. Since this factor was not explicitly controlled for, GenAI adoption rates may be overestimated. Additionally, the timing of the survey (winter term 2024/2025) may have influenced participation, with later-semester students, particularly those in internships, being underrepresented, leading to selection bias. Furthermore, the study did not distinguish between voluntary GenAI adoption and mandatory use, which is essential for understanding students’ true motivations but was not accounted for.

\paragraph{External Validity} The study’s restriction to the case context limits its generalizability, as GenAI policies and educational cultures may differ across institutions. Language barriers may have led to the underrepresentation of international students, reducing the diversity of perspectives. Additionally, disparities in digital literacy and financial resources were not considered, meaning students with limited access to premium GenAI tools may engage with them differently, potentially skewing the findings.

\paragraph{Conclusion Validity} The survey study relied on descriptive statistics from Likert-scale responses, limiting the ability to establish statistical significance or detect subtle GenAI usage patterns. Open-ended responses were analyzed qualitatively, but subjective categorization may have introduced bias, potentially overlooking nuances. Additionally, survey dropout rates could have skewed results, as students more engaged with GenAI or willing to complete surveys may be overrepresented. Without data on dropout reasons, the impact on findings remains unclear.

\section{Conclusion and Future Work}
\label{sec7:Conclusion}
The rapid diffusion of GenAI is fundamentally reshaping academic work practices in higher education. This study examined how students in information systems–related programs engage with GenAI tools and how universities can respond to this development through systematic institutional adaptation. Our findings show that GenAI has already become an integral component of students’ learning workflows, while institutional governance structures are still in the process of adjusting to this technological shift.

Rather than viewing GenAI adoption solely as a pedagogical or technological phenomenon, our study highlights the importance of aligning three institutional dimensions: student practices, academic regulations, and curricular structures. Based on the empirical insights from our case study, we developed the AI Adoption Framework for Higher Education. The framework conceptualizes institutional adaptation as an evidence-informed and iterative process that connects the analysis of existing governance artifacts with empirical observations of student behavior and subsequent institutional adjustments.

The empirical results underline the urgency of such a structured approach. A large proportion of students reported using GenAI tools in their academic work, particularly for tasks such as research support, programming assistance, and conceptual clarification. At the same time, many students expressed uncertainty regarding the permissibility of these practices within existing university regulations. This discrepancy indicates that the challenge of GenAI integration is not limited to technological adoption but also involves issues of transparency, policy communication, and the interpretation of academic integrity norms.

Our findings therefore suggest that universities should move beyond reactive policy adjustments and instead establish mechanisms for continuous institutional learning. The proposed framework provides one possible operationalization of such a mechanism by combining document analysis, empirical data collection, and iterative policy refinement. In doing so, it offers a practical instrument that can support universities in maintaining regulatory clarity while simultaneously adapting curricula and assessment strategies to evolving technological capabilities.

This work also opens several avenues for future research. First, additional case studies across different institutional contexts and academic disciplines are necessary to further evaluate the applicability and robustness of the proposed framework. Comparative studies could provide insights into how variations in governance structures, disciplinary cultures, or institutional policies influence the integration of GenAI in higher education. Second, longitudinal investigations are needed to understand how student usage patterns evolve over time and how institutional responses shape these practices.

Finally, future research should explore the broader educational implications of sustained GenAI use. In particular, it remains important to examine how reliance on AI-supported workflows influences students’ cognitive processes, including critical thinking, problem-solving, and independent reasoning. A deeper understanding of these dynamics will be essential for designing learning environments and assessment strategies that preserve meaningful academic competencies while acknowledging the realities of AI-assisted knowledge production.
%
%
%
 \bibliographystyle{splncs04}
 \bibliography{references}

@article{von_garrel_kunstliche_2023,
	title = {Künstliche Intelligenz im Studium Eine quantitative Befragung von Studierenden zur Nutzung von {ChatGPT} \&amp; Co.},
	rights = {Creative Commons - {CC} {BY} - Namensnennung 4.0 International},
	url = {https://opus4.kobv.de/opus4-h-da/395},
	doi = {10.48444/H_DOCS-PUB-395},
	author = {von Garrel, Jörg and Mayer, Jana and Mühlfeld, Markus},
	urldate = {2024-12-28},
	year = {2023},
	langid = {german}
}

@article{gottschling_nutzung_2024,
	title = {Nutzung von {KI}-Tools durch Studierende},
	issn = {{ISSN}: 2199-8825},
	doi = {DOI: 10.3278/HSL2411W},
	number = {11},
	journaltitle = {die hochschullehre},
	author = {Gottschling, Steffen and Seidl, Tobias and Vonhof, Cornelia},
	year = {2024},
	langid = {german},
}

@INPROCEEDINGS{neumann_we_2023,
  author={Neumann, Michael and Rauschenberger, Maria and Schön, Eva-Maria},
  booktitle={Proc. of the 5th Intnl. Workshop on Software Engineering Education for the Next Generation}, 
  title={“We Need To Talk About ChatGPT”: The Future of AI and Higher Education}, 
  year={2023},
  volume={},
  number={},
  pages={29-32},

  doi={10.1109/SEENG59157.2023.00010}}

@inproceedings{Ruedian.2025,
author = {R\"{u}dian, Sylvio and Podelo, Julia and Ku\v{z}\'{\i}lek, Jakub and Pinkwart, Niels},
title = {Feedback on Feedback: Student’s Perceptions for Feedback from Teachers and Few-Shot LLMs},
year = {2025},
isbn = {9798400707018},
publisher = {Association for Computing Machinery},
address = {New York, NY, USA},
doi = {10.1145/3706468.3706479},
booktitle = {Proceedings of the 15th International Learning Analytics and Knowledge Conference},
pages = {82–92},
numpages = {11},
series = {LAK '25}
}

@ARTICLE{Shailendra.2024,
  author={Shailendra, Samar and Kadel, Rajan and Sharma, Aakanksha},
  journal={IEEE Transactions on Education}, 
  title={Framework for Adoption of Generative Artificial Intelligence (GenAI) in Education}, 
  year={2024},
  volume={67},
  number={5},
  pages={777-785},
  doi={10.1109/TE.2024.3432101}}

@ARTICLE{Chen.2020,
  author={Chen, Lijia and Chen, Pingping and Lin, Zhijian},
  journal={IEEE Access}, 
  title={Artificial Intelligence in Education: A Review}, 
  year={2020},
  volume={8},
  number={},
  pages={75264-75278},
  doi={10.1109/ACCESS.2020.2988510}}

@INPROCEEDINGS{Speth.2023,
  author={Speth, Sandro and Meißner, Niklas and Becker, Steffen},
  booktitle={Proc. of the 35th Intnl. Conf. on Software Engineering Education and Training}, 
  title={Investigating the Use of AI-Generated Exercises for Beginner and Intermediate Programming Courses: A ChatGPT Case Study}, 
  year={2023},
  volume={},
  number={},
  pages={142-146},
  doi={10.1109/CSEET58097.2023.00030 }}

@article{Nithithanatchinnapat.2024,
author = {Nithithanatchinnapat, Benyawarath and Maurer, Joshua and Deng, Xuefei (Nancy) and Joshi, K. D.},
title = {Future Business Workforce: Crafting a Generative AI-Centric Curriculum Today for Tomorrow's Business Education},
year = {2024},
issue_date = {February 2024},
publisher = {Association for Computing Machinery},
address = {New York, NY, USA},
volume = {55},
number = {1},
issn = {0095-0033},
doi = {10.1145/3645057.3645059},
journal = {SIGMIS Database},
month = feb,
pages = {6–11},
numpages = {6}
}

@INPROCEEDINGS{Brockenbrough.2024,
  author={Brockenbrough, Allan and Salinas, Dominic},
  booktitle={Proc. of the 36th Intnl. Conference on Software Engineering Education and Training}, 
  title={Using Generative AI to Create User Stories in the Software Engineering Classroom}, 
  year={2024},
  volume={},
  number={},
  pages={1-5},
  doi={10.1109/CSEET62301.2024.10662994  }}

@INPROCEEDINGS{Speth.2024,
  author={Speth, Sandro and Meißner, Niklas and Becker, Steffen},
  booktitle={Proc. of the 36th Intnl. Conf. on Software Engineering Education and Training}, 
  title={ChatGPT's Aptitude in Utilizing UML Diagrams for Software Engineering Exercise Generation}, 
  year={2024},
  volume={},
  number={},
  pages={1-5},
  doi={10.1109/CSEET62301.2024.10663027 }}

@INPROCEEDINGS{Datta.2024,
  author={Datta, Soma},
  booktitle={Proc. of the 36th Intnl. Conf. on Software Engineering Education and Training}, 
  title={Using Generative Artificial Intelligence Tools in Software Engineering Courses}, 
  year={2024},
  volume={},
  number={},
  pages={1-2},
  doi={10.1109/CSEET62301.2024.10663042 }}

@inproceedings{Savelka.2023,
author = {Savelka, Jaromir and Agarwal, Arav and Bogart, Christopher and Song, Yifan and Sakr, Majd},
title = {Can Generative Pre-trained Transformers (GPT) Pass Assessments in Higher Education Programming Courses?},
year = {2023},
isbn = {9798400701382},
publisher = {Association for Computing Machinery},
address = {New York, NY, USA},
doi = {10.1145/3587102.3588792 },
booktitle = {Proc. of the Conf. on Innovation and Technology in Computer Science Education V. 1},
pages = {117–123},
numpages = {7},
location = {Turku, Finland}
}

@ARTICLE{Schoen.2023, 
AUTHOR={Sch{\"o}n, Eva-Maria  and Neumann, Michael  and Hofmann-St{\"o}lting, Christina  and Baeza-Yates, Ricardo  and Rauschenberger, Maria },
TITLE={How are AI assistants changing higher education?},
JOURNAL={Frontiers in Computer Science},
VOLUME={5},
YEAR={2023},
DOI={10.3389/fcomp.2023.1208550 },
ISSN={2624-9898}
}

@INPROCEEDINGS{Schon.2023b,
author="Eva-Maria Schön and Ilona Buchem and Stefano Sostak and Maria Rauschenberger",
editor="Marchiori, Massimo
and Dom{\'i}nguez Mayo, Francisco Jos{\'e}
and Filipe, Joaquim",
title="Shift Toward Value-Based Learning: Applying Agile Approaches in Higher Education",
booktitle="Web Information Systems and Technologies",
year="2023",
publisher="Springer Nature Switzerland",
address="Cham",
pages="24--41",
isbn="978-3-031-43088-6"
}

@INPROCEEDINGS{Zastudil.2023,
  author={Zastudil, Cynthia and Rogalska, Magdalena and Kapp, Christine and Vaughn, Jennifer and MacNeil, Stephen},
  booktitle={Proc. of the Frontiers in Education Conference}, 
  title={Generative AI in Computing Education: Perspectives of Students and Instructors}, 
  year={2023},
  volume={},
  number={},
  pages={1-9},
  doi={10.1109/FIE58773.2023.10343467 }}

@INPROCEEDINGS{Maher.2023,
  author={Maher, Mary Lou and Tadimalla, Sri Yash and Dhamani, Dhruv},
  booktitle={Proc. of the Frontiers in Education Conference}, 
  title={An Exploratory Study on the Impact of AI tools on the Student Experience in Programming Courses: an Intersectional Analysis Approach}, 
  year={2023},
  volume={},
  number={},
  pages={1-5},
  doi={10.1109/FIE58773.2023.10343037}}

@INPROCEEDINGS{Chan.2023,
  author={Chan, Miguel Morales and Amado-Salvatierra, Hector R. and Hernandez-Rizzardini, Rocael and De La Roca, Mónica},
  booktitle={Proc. of the Frontiers in Education Conference}, 
  title={The potential role of AI-based Chatbots in Engineering Education. Experiences from a teaching perspective}, 
  year={2023},
  volume={},
  number={},
  pages={1-5},
  doi={10.1109/FIE58773.2023.10343296}}

@INPROCEEDINGS{Neumann.2022,
  author={Neumann, Michael and Mötefindt, David and Linke, Lukas and Radtke, Dirk and Mattstädt, Annika and Herzig, Frederik and Regel, Patricia},
  booktitle={Proc. of the 17th Intnl. Conf. on Software Engineering Advances}, 
  title={How to Fill the Gap between Practice and Higher Education: Performing eduScrum with Real World Problems in Virtual Distance Teaching}, 
  year={2022},
  volume={},
  number={},
  pages={1-5}}

@INPROCEEDINGS{Grashoff.2024,
  author={Grashoff, Imke and Mayer, Thomas and Recker, Jan},
  booktitle={ICIS 2024 Proceedings}, 
  title={Challenges, Success Factors, and Potential Benefits of GenAI Development: Insights from an Ongoing Case Study at a German Automotive Manufacturer}, 
  year={2024},
  volume={},
  number={},
  url={https://aisel.aisnet.org/icis2024/aiinbus/aiinbus/20 }}

@INPROCEEDINGS{Matthies.2022,
  author={Matthies, Christoph  and Teusner, Ralf and Perscheid, Michael},
  booktitle={Proc. of the 55th Hawaii Intnl. Conf. on System Sciences}, 
  title={Challenges (and Opportunities!) of a Remote Agile Software Engineering Project Course During COVID-19}, 
  year={2022},
  volume={},
  number={},
  url={http://hdl.handle.net/10125/79444}}

@InProceedings{Sami.2025, 
author="Sami, Malik Abdul
and Waseem, Muhammad
and Zhang, Zheying
and Rasheed, Zeeshan
and Syst{\"a}, Kari
and Abrahamsson, Pekka",
editor="Pfahl, Dietmar
and Gonzalez Huerta, Javier
and Kl{\"u}nder, Jil
and Anwar, Hina",
title="Early Results of an AI Multiagent System for Requirements Elicitation and Analysis",
booktitle="Product-Focused Software Process Improvement",
year="2025",
publisher="Springer Nature Switzerland",
address="Cham",
pages="307--316",
isbn="978-3-031-78386-9"
}

@InProceedings{Rasheed.2024,
author="Rasheed, Zeeshan
and Waseem, Muhammad
and Sami, Malik Abdul
and Kemell, Kai-Kristian
and Ahmad, Aakash
and Duc, Anh Nguyen
and Syst{\"a}, Kari
and Abrahamsson, Pekka",
editor="Marchesi, Lodovica
and Goldman, Alfredo
and Lunesu, Maria Ilaria
and Przyby{\l}ek, Adam
and Aguiar, Ademar
and Morgan, Lorraine
and Wang, Xiaofeng
and Pinna, Andrea",
title="Autonomous Agents in Software Development: A Vision Paper",
booktitle="Agile Processes in Software Engineering and Extreme Programming -- Workshops",
year="2025",
publisher="Springer Nature Switzerland",
address="Cham",
pages="15--23",
isbn="978-3-031-72781-8"
}

@InProceedings{Zhang.2024,
author="Zhang, Zheying
and Rayhan, Maruf
and Herda, Tomas
and Goisauf, Manuel
and Abrahamsson, Pekka",
title="LLM-Based Agents for Automating the Enhancement of User Story Quality: An Early Report",
booktitle="Agile Processes in Software Engineering and Extreme Programming",
year="2024",
publisher="Springer Nature Switzerland",
address="Cham",
pages="117--126",
isbn="978-3-031-61154-4"
}

@article{Runeson.2009,
 author = {Runeson, P. and H{\"o}st, M.},
 year = {2009},
 title = {Guidelines for conducting and reporting case study research in software engineering},
 pages = {131--164},
 volume = {14},
 number = {2},
 issn = {1382-3256},
 journal = {Empirical Software Engineering},
 doi = {\url{10.1007/s10664-008-9102-8}}
}

@book{Yin.2009,
 author = {Yin, Robert K.},
 year = {2009},
 title = {Case study research: Design and methods},
 OPTaddress = {Los Angeles},
 edition = {4th},
 volume = {5},
 publisher = {Sage},
 isbn = {9781412960991},
 series = {Applied social research methods series}
}

@article{Lederman.2015,
 author = {Lederman, N.G. and Lederman, J.S.},
 year = {2015},
 title = {What Is A Theoretical Framework? A Practical Answer},
 pages = {593–-597},
 volume = {26},
 number = {},
 journal = {J Sci Teacher Educ},
 doi = {\url{10.1007/s10972-015-9443-2  }}
}

@INPROCEEDINGS{Cruzes.2011,
  author={D. S. Cruzes and T. Dyba},
  booktitle={Proc. of the Intnl. Symposium on Empirical Software Engineering and Measurement}, 
  title={Recommended steps for thematic synthesis in software engineering}, 
  year={2011},
  volume={},
  publisher={IEEE},
 pages = {275–284},
  url={http://hdl.handle.net/10125/79444}}

@INPROCEEDINGS{Foerster.2024,
  author={Maximilian Förster and Kirsten Pitz and Andrea Wrabel and Mathias Klier and
Steffen Zimmermann},
  booktitle={ Wirtschaftsinformatik 2024 Proceedings}, 
  title={Building AI Literacy with Experiential Learning – Insights from a Field Experiment in K-12 Education}, 
  year={2024},
  url={https://aisel.aisnet.org/wi2024/24 }}

@article{Southworth.2023,
title = {Developing a model for AI Across the curriculum: Transforming the higher education landscape via innovation in AI literacy},
journal = {Computers and Education: Artificial Intelligence},
volume = {4},
pages = {100127},
year = {2023},
issn = {2666-920X},
doi = {https://doi.org/10.1016/j.ca          eai.2023.100127}         ,
author = {Jane Southworth and Kati Migliaccio and Joe Glover and Ja’Net Glover and David Reed and Christopher McCarty and Joel Brendemuhl and Aaron Thomas}
}

@article{Su.2023,
 author = {J. Su and W. Yang},
 year = {2023},
 title = {Unlocking the power of ChatGPT: A framework for applying generative AI in education},
 pages = {355–366},
 volume = {6},
 number = {3},
 journal = {ECNU Rev. Educ.}
}

@misc{Bischof.2024,
author = {Lasse Bischof and Maria Rauschenberger and Eva-Maria Schön and Michael Neumann},
year = {2024},
url = {https://doi.org/10.5281/zenodo.15017474                                                    },
title = {Questionnaire GenAI usage by students}
}

@INPROCEEDINGS{Freise.2025,
  author={Leonie Rebecca Freise  and Olivia Bruhin and Eva Ritz and Mahei Manhai Li and Jan Marco Leimeister},
  booktitle={Proc. of the 58th Hawaii Intnl. Conf. on System Sciences}, 
  title={Code and Craft: How Generative AI Tools Facilitate Job Crafting in Software Development}, 
  year={2025},
  volume={},
  number={},
  url={https://hdl.handle.net/10125/109683}}

@inproceedings{Jimenez.2024,
title={{SWE}-bench: Can Language Models Resolve Real-world Github Issues?},
author={Carlos E Jimenez and John Yang and Alexander Wettig and Shunyu Yao and Kexin Pei and Ofir Press and Karthik R Narasimhan},
booktitle={Proc. of the 12th Intnl. Conf. on Learning Representations},
year={2024}
}

@conference{Bischof.2025,
author={Lasse Bischof and Eva-Maria Schön and Maria Rauschenberger and Michael Neumann},
title={“We Need to Analyze Students GenAI Use”: Towards an AI Adoption Framework for Higher Education},
booktitle={Proceedings of the 21st International Conference on Web Information Systems and Technologies},
year={2025},
pages={429-438},
publisher={SciTePress},
organization={INSTICC},
doi={10.5220/0013819000003985},
isbn={978-989-758-772-6},
}

@article{Aldossary.2024,
 author = {Aldossary, A. S. and Aljindi, A. A. and Alamri, J. M.},
 year = {2015},
 title = {The role of generative AI in education: Perceptions of Saudi students},
 pages = {},
 volume = {16},
 number = {4},
 journal = {Contemporary Educational Technology},
 doi = {\url{10.30935/cedtech/15496}}
}

@article{Mah.2024,
  author       = {Mah, D. K. and Groß, N.},
  title        = {Artificial intelligence in higher education: exploring faculty use, self-efficacy, distinct profiles, and professional development needs},
  journal      = {International Journal of Educational Technology in Higher Education},
  year         = {2024},
  volume       = {21},
  pages        = {58},
  doi          = {10.1186/s41239-024-00490-1}
}

@article{Wang.2025,
  author       = {Wang, C.},
  title        = {Exploring Students’ Generative AI-Assisted Writing Processes: Perceptions and Experiences from Native and Nonnative English Speakers},
  journal      = {Technology, Knowledge and Learning},
  year         = {2025},
  volume       = {30},
  pages        = {1825--1846},
  doi          = {10.1007/s10758-024-09744-3},
  url          = {https://doi.org/10.1007/s10758-024-09744-3}
}

@article{Rizki.2025,
  author       = {Rizki, I.~A. and Daoud, R.},
  title        = {Generative Artificial Intelligence in Higher Education: Review of Institutional Policies and Practices across New Zealand},
  journal      = {New Zealand Journal of Educational Studies},
  year         = {2025},
  doi          = {10.1007/s40841-025-00417-y},
  url          = {https://doi.org/10.1007/s40841-025-00417-y}
}

@article{Wiese.2025,
title = {AI ethics education: A systematic literature review},
journal = {Computers and Education: Artificial Intelligence},
volume = {8},
pages = {100405},
year = {2025},
issn = {2666-920X},
doi = {https://doi.org/10.1016/j.caeai.2025.100405},
url = {https://www.sciencedirect.com/science/article/pii/S2666920X25000451},
author = {Lucas J. Wiese and Indira Patil and Daniel S. Schiff and Alejandra J. Magana},
}
 \end{document}